\documentclass[aps,prf,onecolumn,citesort]{revtex4-2}

\usepackage{graphicx}
\usepackage{dcolumn}
\usepackage{bm}
\usepackage{xfrac}
\usepackage{siunitx}
\usepackage{amsmath}
\usepackage{soul}

\begin{document}


\preprint{APS/123-QED}

\title{
Vertical transport and confinement of weakly buoyant particles\\ in a convective ocean mixed-layer model}

\author{Luz Andrea Silva-Torres}
\email{Contact author: luz-andrea.silva-torres@univ-lille.fr}
\affiliation{
Univ. Lille, ULR 7512–Unité de Mécanique de Lille Joseph Boussinesq (UML), F-59000 Lille, France
}
\author{Enrico Calzavarini}
\affiliation{ 
Univ. Lille, ULR 7512–Unité de Mécanique de Lille Joseph Boussinesq (UML), F-59000 Lille, France
}
\author{Stefano Berti}
\affiliation{
Univ. Lille, ULR 7512–Unité de Mécanique de Lille Joseph Boussinesq (UML), F-59000 Lille, France
}%



\date{\today}

\begin{abstract}

Understanding how particles such as microplastics and plankton spread, or accumulate in localized regions below the ocean surface is crucial both for the development of sea pollutant control strategies and for marine ecology. Yet the role of vertically inhomogeneous turbulence, due to the vertical structure of density stratification in the ocean, remains poorly understood. We study, by means of direct numerical simulations, the dynamics of weakly inertial, quasi-neutrally buoyant particles in an idealized two-layer fluid model aimed to represent the ocean convective mixed layer and the more stably stratified upper thermocline beneath. We find that particles eventually rise at the surface if their density matches the mean value of the fluid at a selected depth within the mixed layer. Interestingly, however, for all other values of density, they accumulate around their neutral-buoyancy depth within the thermocline, where stratification hinders vertical transport. Our analysis shows that the thickness of the accumulation layer results from the competition between buoyancy-driven confinement, internal-wave motions, and turbulent fluctuations penetrating from the mixed layer. Smaller particles, which respond more rapidly to flow fluctuations, form broader layers; layers also broaden as their equilibrium depth approaches the mixed layer. We then derive a reduced stochastic model that explains the dependence  on particle inertia and reference depth of such spreading around the equilibrium position.  These results show how the coupled action of stratification and decaying mixed-layer turbulence controls subsurface particle-layer formation, which may be relevant to ocean-biology oriented studies and to improve the sampling of plastic pollution in the sea.

\end{abstract}

\maketitle


\section{Introduction}\label{sec1}
In the environment, stably stratified flows are ubiquitous, occurring in the ocean, the atmosphere, estuaries, and lakes. Such flows are characterized by the increase of density with depth and support the propagation of internal gravity waves. Under stable stratification, buoyancy acts as a restoring force against vertical displacements, thereby hindering vertical motions. In the ocean, this stratification gives rise to a layered vertical structure, with a well-mixed surface layer overlying the strongly stratified thermocline and the 
less stratified abyss~\cite{Sutherland_2010}. Under external forcing, internal gravity waves can become unstable and break, providing a source of turbulence. Turbulence can also be generated through interactions with the atmosphere, for instance by wind-driven mixing and thermal convection. As a result, oceanic flows exhibit a nontrivial interplay between internal gravity waves and turbulent motions spanning a broad range of spatial and temporal scales. These motions play a crucial role in vertical transport, influencing the distribution of biogeochemical tracers and pollutants~\cite{waite2004stratified,armenio2007environmental}.

Understanding how these different dynamical processes transport and redistribute inertial particles is therefore essential for predicting the fate of suspended particulate material of natural and anthropogenic origin in the marine environment. Over the past two decades, increasing attention has been devoted to marine plastic pollution, in particular to determine possible pathways and accumulation hotspots of plastic debris both at the ocean surface and at depth, owing to its severe impacts on ocean health~\cite{Sutherland_2010, borrelle2020predicted}. The task, however, remains challenging because of the different interactions between physical, chemical, and biological processes acting across a broad range of spatial and temporal scales~\cite{andrady2011microplastics, van2020physical}. Similarly complex, multiscale processes also shape the distribution of plankton in the sea, which is typically highly heterogeneous and characterized by localized patches and thin layers~\cite{durham2012thin}.

Considerable research on particle-laden turbulent flows has been conducted in homogeneous, isotropic, and statistically stationary turbulence, where a range of fundamental particle–flow interaction mechanisms have been identified. In particular, inertial clustering, preferential sampling, and particle segregation have been extensively characterized through numerical simulations (see, e.g.,~\cite{calzavarini2008dimensionality, CalzavariniPRLsegregation2008, BecGustavssonMehligARFM2024}). Finite-size and finite-Reynolds number effects on the dynamics of spherical bubbles and solid particles have also been considered ~\cite{calzavarini2009acceleration,zhang2021fluid,brandt2022particle}, as well as the feedback effects of particles on the carrier fluid~\citep{tanaka2008classification, gore1989effect, ferrante2003physical}.

Although this idealized framework is valuable for elucidating fundamental physical mechanisms, its assumptions of homogeneity and isotropy are only approximately satisfied locally at sufficiently small scales in the ocean. At larger scales, inhomogeneity and anisotropy become increasingly important. A natural first extension of this framework, particularly for problems involving vertical transport, is therefore to account for stable stratification, which introduces a preferred direction and fundamentally alters turbulent transport. 
Early studies revealed the strong suppression of vertical dispersion by buoyancy~\cite{csanady1964turbulent, pearson1983statistical}. Indeed, fluid particles (i.e., Lagrangian tracers) do not diffuse away from their initial positions. In the case of moderate to strong stratification, the mean-square vertical displacement ceases to grow after a time comparable with the inverse of the (Brunt-V\"ais\"al\"a) buoyancy frequency $N$ and approaches an asymptotic value proportional to the ratio of the mean-square vertical velocity and $N^2$.
This suppression of vertical dispersion is closely related to the behavior of the Lagrangian pressure-gradient 
acceleration, which contributes to the decorrelation of the velocity field~\cite{nicolleau2000turbulent}. At sufficiently long times, however, vertical dispersion can recover a diffusive regime if turbulence is strong enough~\cite{csanady1964turbulent, pearson1983statistical, nicolleau2000turbulent}. 

Moreover, it has been shown that while horizontal fluid-particle dispersion remains qualitatively similar to that observed in homogeneous and isotropic turbulence, vertical dispersion exhibits a markedly different behavior linked to the strength of the background stratification~\cite{van2008single}. At short times, internal gravity waves dominate the particle vertical dynamics, giving rise to an initial ballistic regime. At intermediate times, the interplay between wavy and turbulent motions can lead to distinct transient behaviors, which may or may not eventually give way to the asymptotic diffusive regime. At late times, the saturation of single-particle vertical dispersion appears to depend on both the nature of the forcing and the competition between turbulent mixing and stratification~\cite{sujovolsky2019vertical}. Given the prominent role of internal gravity waves, models based on the turbulent vertical Lagrangian velocity spectrum have been proposed to predict the vertical dispersion of tracers~\cite{sujovolsky2018single}. Recently, laboratory experiments have provided further evidence for the saturation of vertical dispersion~\cite{magnier2026lagrangian}. Numerical simulations have also revealed that intermittent extreme events in stratified turbulence can strongly influence vertical particle separation, producing pronounced anisotropy in their dispersion and enhanced vertical mixing in localized regions~\cite{reartes2026extreme}.

Regarding inertial particles, direct numerical simulations (DNS) in stratified turbulence have highlighted the departure of particle dynamics from that of fluid tracers as a function of Stokes number (which quantifies the relevance of particle inertia). Stratification still tends to suppress vertical motions, but the resulting confinement can now be significantly affected by inertial effects~\cite{van2010vertical}. The relative importance of the different forces acting on particles was also investigated numerically in turbulent flows of geophysical interest~\cite{van2010vertical,reartes2024bounds}. It was recently shown, in particular, that the Basset–Boussinesq history force can often be neglected in most oceanic situations, as the latter becomes significant only for very strong stratification or for particles with sufficiently large inertia~\cite{reartes2024bounds}.

Motivated by observations of marine particles accumulating in, often quite thin, layers below the ocean surface~\cite{durham2012thin,uurasjarvi2021microplastics,zhao2025distribution}, a number of numerical studies have investigated the mechanisms responsible for such vertical confinement and the factors controlling the thickness of these layers. The vertical distribution of floating particles, as those considered in the present work, was first studied in terms of passive particles subjected to a linear restoring force producing vertical confinement, in three-dimensional (3D) homogeneous, isotropic turbulence~\cite{de2015clustering}. 
It was found, in particular, that particles experience, in this case, an effective compressibility and accumulate over fractal clusters. 
Adopting a similar approach, the analysis was subsequently refined to explicitly account for stratification and to introduce a more physically justified model, based on the Boussinesq approximation, for the dynamics of small buoyant particles~\cite{sozza2016large}. The results showed that vertical confinement is governed primarily by the strength of the background stratification and is accompanied by pronounced fractal clustering over isopycnal surfaces.  
A further development of this study demonstrated that the thickness of these layers, where particles accumulate, results from the competition between buoyancy and turbulence, with particle inertia also playing a central role. This leads to a non-monotonic dependence of layer thickness on particle properties, which can be captured by a stochastic model of vertical transport~\cite{sozza2018inertial}. 
More recent work has shown, in addition, that such particles behave as damped oscillators forced by internal gravity waves~\cite{reartes2023dynamical} and that depending on both particle inertia (Stokes number) and stratification strength (Brunt–V\"ais\"al\"a frequency), their response can be overdamped or oscillatory. Stratification confines their vertical dispersion, while particle inertia changes how they respond to fast vertical motions.

Taken together, these studies have established several fundamental aspects of the dynamics of passive, inertial particles in stratified flows. However, they have focused on idealized configurations consisting of a single homogeneous turbulent layer.
In the present work, in the spirit of adding a further essential feature of oceanic flows, we consider a system made of two layers with different stratification, which can be seen as a simplified representation of both the surface mixed layer and the underlying upper thermocline. 
For this purpose, we adopt a two-dimensional (2D), convective mixed-layer model~\cite{bhamidipati2020turbulent} and explore the dynamics of quasi-neutrally buoyant particles in the resulting flows by means of DNS. 
We particularly investigate the vertical particle distribution in the thermocline, focusing on the role of turbulence, whose intensity decays with depth below the mixed layer. Also in this case, particles undergo vertical confinement as a result of stable stratification. Moreover, generally speaking, larger particles get more confined than smaller ones, in the range of parameters explored. However, we find that the extent of the particle accumulation regions is strongly influenced by turbulent fluctuations originating in the mixed layer and penetrating into the thermocline. 
In addition, the interplay between stratification and vertically inhomogeneous turbulence produces quite complex Lagrangian behaviors, due to the different types of motion experienced by particles over  
different temporal scales. To interpret the observed dynamics, we propose a simple stochastic, wave-driven relaxation model that incorporates the effects of internal gravity waves and turbulent fluctuations on vertical particle transport.

This article is organized as follows. In Sec.~\ref{sec2} we describe the models adopted for both the two-layer, stratified flow (Sec.~\ref{sec:model_eul}) and the dynamics of the quasi-neutrally buoyant particles (Sec.~\ref{sec:model_lagr}), and provide a summary of the numerical methods and parameters employed for the simulations (Sec.~\ref{sec:num_sim}). Section~\ref{sec3} reports the results. After illustrating the phenomenology for the possible cases, with respect to the reference density at which particles are neutrally buoyant, and for varying inertia, we turn to the analysis of the vertical particle distributions. In particular, we focus on the phenomenon of particle accumulation and quantify the size of the layer over which the latter takes place. We then derive the stochastic model and use it to explain these findings. Finally, discussions and conclusions are presented in Sec.~\ref{sec4}.

\section{\label{sec2}Model and Numerical Simulations}

\subsection{Eulerian Dynamics: A Convective Mixed Layer Model} 
\label{sec:model_eul}

We consider a vertically inhomogeneous, stratified fluid, intended to reproduce the qualitative structure of the upper ocean. In particular, we aim to account for the presence of a weakly stratified, upper mixed layer, essentially homogenized by turbulent motions, on top of a deeper layer characterized by a much stronger stable stratification (constant with depth, for simplicity), akin to the upper portion of the ocean thermocline. 
For this purpose, we adopt the idealized 2D convective mixed-layer model first introduced in \cite{bhamidipati2020turbulent}, in which the surface buoyancy flux, resulting from cooling at the air-sea interface, is the driver of mixing. A balance between surface buoyancy flux and internal stratification is maintained long enough for the system to reach a statistically steady state. 
The mathematical description of the system relies on the Boussinesq equations with an internal heating term in the temperature (and, hence, buoyancy) equation due to penetrative solar radiation. This configuration enables surface cooling, generating buoyancy-driven instabilities that promote convective mixing. The governing equations for fluid dynamics in space ($X,Z$) an time ($\mathcal{T}$) are:
\begin{subequations}
\label{e1}
\begin{align}
& \ \nabla\cdot\bm{U} = 0, \label{e1a} \\
& \ \frac{\partial \bm{U}}{\partial \mathcal{T}} + (\bm{U}\cdot \nabla) \bm{U} = -\frac{1}{\rho_0} \nabla P^\ast +B \hat{\bm{Z}}+\nu\nabla^2\bm{U}, \label{e1b} \\
& \ \frac{\partial B}{\partial \mathcal{T}} + \bm{U}\cdot\nabla B= \kappa\nabla^2 B + \frac{dQ({Z})}{d{Z}},
\label{e1c}
\end{align}
\end{subequations}
where $\bm{U}(X,Z,\mathcal{T}) = (U_X, U_Z)$ is the two-dimensional velocity field, $\nu$ is the kinematic viscosity, and $\kappa$ is the thermal diffusivity. The modified pressure is defined as $P^\ast = P + \alpha \rho_0 T_0 g Z$, with $P$ denoting  
pressure fluctuations, $g$ gravity acceleration, and $\rho_0$ and $T_0$ the reference density and temperature, respectively. Note that here we assume that density changes due to pressure compressibility and haline effects are much smaller than those due to thermal expansion \citep{vallis2017atmospheric}. Under these approximations, the fluid density $\rho_f$ at any point can be expressed as $\rho_f = \rho_0[1 - \alpha(T - T_0)]$, with $\alpha$ the thermal expansion coefficient, and buoyancy $B=\alpha g T$ only depends on temperature. The internal heating term is given by the Beer-Lambert law $Q({z})=Q_0 e^{{Z}/l}$, where $-H \leq Z \leq 0$, with $Z=0$ and $Z=-H$ respectively indicating the sea surface and the largest considered depth; $l$ is the attenuation length of incoming short-wave solar radiation penetrating the water column, and $Q_0=H_0(\alpha g)/(\rho_0 c_p)$. Here, $c_p$ is the specific heat capacity of seawater, and $H_0$ is the surface heat flux, which is assumed to be constant for simplicity. In reality, however, $H_0$ varies diurnally and seasonally, and is composed of the latent, sensible, short-wave, and long-wave components \citep{josey2023declining}.

As for boundary conditions, we assume free-slip conditions for the velocity field at the top and bottom boundaries. Concerning the buoyancy field, following~\cite{bhamidipati2020turbulent}, a bottom fixed heat flux is imposed to sustain the background stratification and preserve the thermocline structure over time. The Brunt–Väisälä frequency $N$ varies with depth, but we verified that the temperature vertical profile remains essentially linear and steady during our simulations. The surface boundary condition for buoyancy is derived by performing a volume average, denoted by $\langle \cdot \rangle$, of Eq.~(\ref{e1c}), and applying the divergence theorem. A global steady state is assumed; that is $\partial_\mathcal{T} \langle B \rangle = 0$. This boundary condition is implemented locally in the model, allowing the system to reach a statistically steady state. In summary, the boundary conditions for Eqs.~(\ref{e1b}) and~(\ref{e1c}) are:
\begin{subequations}
\label{e2}
\begin{align}
\text{at  $Z = -H$}: \quad & U_Z = 0, \quad \frac{\partial U_X}{\partial Z} = 0, \quad 
\kappa \frac{\partial B}{\partial Z} = \kappa N^2, \label{2a} \\
\text{at  $Z = 0$}: \quad & U_Z = 0, \quad \frac{\partial U_X}{\partial Z} = 0,
\quad  \kappa \frac{\partial \langle B \rangle_{X}}{\partial Z} = \kappa   N^2 - Q_0 \left( 1 - e^{-H/l}
\right),
 \label{2b}
\end{align}
\end{subequations}
where $\langle \cdot \rangle_X$ stands for an average on the horizontal. We note that when the system domain is sufficiently deep with respect to the radiative attenuation length $l$ -~as it will be the case in our simulations~- the term with the exponential can be neglected. As reported in \cite{bhamidipati2020turbulent}, to make convection possible, the slope of the buoyancy field at the surface must satisfy $\partial_Z B < 0$, so that colder (heavier) fluid lies above warmer (lighter) fluid. 

It is useful to recast the governing equations in nondimensional form. For this, we introduce the following nondimensional variables: 
\begin{equation}
t= \mathcal{T}N, \quad \bm{u} = \frac{\bm{U}}{(Q_0l)^{1/3}}, \quad \bm{x}= \frac{\bm{X}N}{(Q_0l)^{1/3}}, \quad b = \frac{B}{(Q_0l)^{1/3}N}, \quad p^* = \frac{P^*}{(Q_0l)^{2/3}\rho_0}.
\label{eq:nondim_var}
\end{equation}

Here, our choices for the nondimensional variables differ from those of~\cite{bhamidipati2020turbulent}, since we aim to provide a suitable representation of the particle dynamics in the different regions of the flow, particularly in the thermocline, where turbulent motions originating from the layer above coexist with strong fluid stratification. Specifically, we select the characteristic timescale associated with stratification in the thermocline, which is given by $1/N$, and the characteristic velocity scale generated in the mixed layer through the radiative forcing, given by $(Q_0 l)^{1/3}$. The remaining nondimensional variables are then derived from these two primary ones. With these definitions, Eqs.~(\ref{e1a})-(\ref{e1c}) become:

\begin{subequations}
\label{e3}
\begin{align}
& \nabla\cdot\bm{u} = 0 \label{e3a}\\
& \dfrac{\partial\bm{u}}{\partial t} + \left( \bm{u} \cdot \nabla \right)\bm{u} = -\nabla p^*+ b \hat{\bm{z}} + \dfrac{1}{Re} \nabla^2\bm{u}\label{e3b} \\
& \dfrac{\partial b}{\partial t} +  \bm{u}\cdot \nabla b = \dfrac{1}{Pe} \nabla^2 b+\Phi^2 e^{\Phi z}.
\label{e3c}
\end{align}
\end{subequations}
The nondimensional control parameters are then the following ones:
\begin{equation}
Re = (Q_0l)^{2/3}/\nu N, \quad 
Pe=(Q_0l)^{2/3}/\kappa N, \quad 
\Phi = Q_0^{1/3}/(l^{2/3}N),
\end{equation}
respectively, the Reynolds number and  
the Péclet number (whose ratio gives the Prandtl number $\mathrm{Pr}=Pe/Re$), and the buoyancy source term, $\Phi$,  expressing the ratio between the destabilizing radiative heating and the stabilizing effect of the bottom pycnocline.

The nondimensional equations are then subjected to the boundary conditions [from~(\ref{2a})-(\ref{2b})]:
\begin{subequations}
\label{e4}
\begin{align}
\label{e4a}
\text{at  $z = -h$}: \quad & u_z = 0, \quad \frac{\partial u_x}{\partial 
z} = 0, \quad 
\frac{\partial b}{\partial z} = 1 \\
\label{e4b}
\text{at  $z = 0$}: \quad & u_z = 0, \quad \frac{\partial u_x}{\partial z} = 0,
\quad \frac{\partial\langle  b\rangle_x}{\partial z} = 1 - Pe\ \Phi  \left( 1 - e^{-\Phi h} \right).
\end{align}
\end{subequations}
Note that for a finite domain there is indeed a fourth dimensionless parameter, $h$ (with $h$ the nondimensional domain depth). Its effect can be neglected in the limit $h\to \infty$ as one can understand from Eq.~(\ref{e4b}). Furthermore, integrating the latter term in Eq.~(\ref{e3c}) over the domain depth, we obtain $\Phi$, which allows us to better grasp the meaning of this parameter as the global dimensionless intensity of the radiative heating.

\subsection{Lagrangian Dynamics: Modeling Quasi-Neutrally Buoyant Particles}
\label{sec:model_lagr}

To accurately capture the dynamics of marine particles, it is necessary to account for particle inertia. For this purpose, one typically resorts to the Maxey-Riley-Gatignol (MRG) equation \citep{maxey1983equation,gatignol1983faxen,michaelides2003hydrodynamic,cartwright2010dynamics}, which incorporates the effects of pressure, buoyancy, Stokes drag, added mass, and the Basset-Boussinesq (history) forces, in addition to Fax\'en corrections that account for the non-uniformity of the flow at the particle scale. 

In the case of interest for this study, particle sizes and velocities are assumed to be small enough to ensure a particle Reynolds number smaller than one, $Re_p = \vert \bm{V} - \bm{U}\vert a/\nu \ll 1$, where 
$\bm{V}(\mathcal{T})=(V_X,V_Z)$ is the two-dimensional Lagrangian particle velocity and $a$ is the particle radius. This condition is essential for the validity of the MRG equation, and moreover, justifies neglecting both Fax\'en corrections and the Basset-Boussinesq force, since their contributions become negligible in the small-particle limit \citep{calzavarini2009acceleration,reartes2023dynamical,monroy2017modeling}. Based on these assumptions, a particle trajectory, $\bm{R}(\mathcal{T})$, is described by the following equations:
\begin{subequations}
\label{e5}
\begin{align}
& \ \bm{\dot{R}} = \bm{V}, \label{e5a}\\
& \  \bm{\dot{V}}= \beta 
\frac{D\bm{U}}{D \mathcal{T}} + 
\frac{\bm{U}-\bm{V}}{\tau_p} + (1 - \beta)\bm{g},
\label{e5b}
\end{align}
\end{subequations}
where the dot-notation is used for the Lagrangian derivative, $\beta = {3\rho_f}/{(2\rho_p+\rho_f)}$, and $\tau_p = a^2/(3\beta\nu)$ is the particle relaxation time. Here, $\rho_f$ and $\rho_p$ denote the fluid and particle densities, respectively. If we consider the limit of small but non-zero particle inertia, i.e., the so called overdamped limit~\cite{maxey1983equation,de2015clustering, sozza2016large,sozza2018inertial}, a perturbative solution of Eq.~\eqref{e5b} yields the approximate particle velocity:

\begin{equation}
    \bm{V} \approx \bm{U}+\tau_p(1-\beta)\left( \bm{g} - \dfrac{D\bm{U}}{D \mathcal{T}}\right).
    \label{e6}
\end{equation}

In our carrier fluid, density varies with depth. To account for this, we consider particles that are quasi-neutrally buoyant. In particular, we assume that the particle density matches the fluid density at a reference depth $z_0$, i.e., $\rho_p = \rho_f(z_0)$. Fractional density changes in seawater are typically very small. The Boussinesq approximation exploits this property \citep{vallis2017atmospheric}, making it applicable to both the fluid and particle dynamics \cite{de2015clustering, sozza2016large, sozza2018inertial}. Using a Boussinesq-type approximation, we then assume $\rho_p \simeq \rho_f(z_0) \simeq \rho_0$. However, density fluctuations remain relevant in the term $(1-\beta)$ because it is multiplied by gravity acceleration. To retain this contribution, we expand $\beta$ around temperature at $z_0$, using the relation $\rho_f \simeq \rho_0[1 - \alpha (T - T_{z_0})]$. This leads to the approximation $\beta \simeq 1 - (2/3) \alpha (T-T_{z_0})$. Following this reasoning, Eq.~\eqref{e6}, becomes:
\begin{equation}
\bm{V} \approx \bm{U}
- \frac{2}{3}\tau_p (B-B_{Z_0})
\left(\hat{\bm{Z}} + \frac{1}{g}\frac{D\bm{U}}{D \mathcal{T}}\right).
\label{e7}
\end{equation}
Recall that here $B_{Z_0}$ denotes the buoyancy (i.e., temperature) at the isopycnal surface $\rho_f=\rho_f(z_0)$, with respect to which the particles are assumed to be neutrally buoyant. Note, moreover, that the particle response time is $\tau_p \simeq a^2 / (3\nu)$, using $\beta \simeq 1$.  

Using the expressions in Eq.~(\ref{eq:nondim_var})  the particle velocity can be written in nondimensional form as
\begin{equation}
\bm{v}= \bm{u}-\dfrac{2}{3}
St( b- b_{ z_0})\left( \hat{\bm{z}}+
Fr\dfrac{D\bm{u}}{D t}\right),
\label{e8}
\end{equation}
where ${D}/{D}{t}=\partial_{{t}} + {\bm{u}} \cdot {\bm{\nabla}}$, $St = \tau_p N$ is the Stokes number, and $Fr = (Q_0l)^{1/3}N/g$ the Froude number. Here, the latter control parameter arises from the choice of nondimensional variables and quantifies the relative importance of inertial accelerations to buoyancy effects in the particle equation. As it will be shown below, the value of $Fr$ in our simulations is sufficiently small to justify neglecting the contribution of the inertial term $D\bm{u}/ Dt$ in the analysis of particle dynamics.

This reduced model (see also \cite{de2015clustering, sozza2016large, sozza2018inertial}) represents the simplest formulation commonly used to describe the dynamics of neutrally buoyant small particles at a reference depth in stably stratified turbulence. The presence of a mixed layer complicates the mechanisms governing particle transport and distribution with respect to the situations investigated in previous studies, but allows for qualitative features of the flow that better resemble those found in real oceanic conditions.

\subsection{Numerical simulations}
\label{sec:num_sim}
We perform DNS of the dynamics from the flow and particle models detailed in Sec.~\ref{sec:model_eul} and Sec.~\ref{sec:model_lagr}, using the \texttt{ch4-project} code \citep{calzavarini2019eulerian}. The flow is solved using the Lattice Boltzmann (LB) method based on the standard stream-and-collide algorithm with a single-relaxation-time collision operator. Two distribution functions are employed, one for the fluid velocity field and one for the buoyancy (or temperature) field, both defined on a D2Q9 lattice. The computational domain is discretized using a uniform Cartesian grid \citep{succi2018lattice,calzavarini2019eulerian}. 
  
Particles are one-way coupled to the flow, so that they do not affect the velocity, density, or temperature fields. The flow velocity at particle positions is obtained via bilinear interpolation, and particle trajectories are integrated in time using a second-order Adams–Bashforth scheme.

The simulations were performed in a 2D square domain with linear size $L_{x}=L_{ z}=19.3$ at resolution $N_x=N_z=400$ and adopting periodic boundary conditions along the horizontal direction. We consider a Prandtl number $Pr=7.2$, which appears reasonable for seawater. The Reynolds and Péclet numbers of the simulated flows are $Re=32$ and $Pe=231$, respectively. The nondimensional buoyancy source parameter is $\Phi=0.51$, which together with $h=19.3$ gives $e^{- \Phi h}=4.5 \cdot 10^{-5}\ll1$ in Eq.~(\ref{e4b}).While these values are clearly not comparable to actual typical oceanic values, we stress that our aim is not to carry out a fully resolved DNS with realistic parameters, which is clearly unfeasible due to the very broad range of scales involved. Rather, our goal is to have at our disposal a flow possessing some of the key qualitative features that characterize upper-ocean flows, to investigate the basic mechanisms controlling the vertical particle distribution. In this sense, in our view the most important aspect concerns the flow nonhomogeneity along the vertical, with a turbulent and well-mixed region lying on top of a more stratified and progressively (with depth) more quiescent one.

We further note that the buoyancy Reynolds number $\mathrm{Re}_b={\varepsilon}/{(\nu N^2)}$, with $\varepsilon$ the kinetic energy dissipation rate, varies significantly with depth, ranging from $Re_b \approx 33$ to $\approx 4.2 \cdot 10^{-2}$ (at depth). This reflects the coexistence of weakly stratified turbulent regions and strongly stratified ones. The Kolmogorov dissipative length scale $\eta=\left(\nu^3/\varepsilon\right)^{1/4}$ remains larger than the grid spacing throughout the turbulent mixed layer, ensuring that small flow scales are always well resolved. 

To investigate the dynamics of particles of different sizes, we considered 
15 different particle types, each characterized by a different diameter, and hence a corresponding response time $\tau_p$, with Stokes numbers ranging from $St = 2.86 \cdot 10^{-6}$ to $9.94 \cdot 10^{-2}$, equivalent to particle diameters between 0.1 \si{\milli \meter} to 22 \si{\milli \meter} in water. This choice allows us to explore the role of particle inertia on material transport in such a convective, stratified flow. For each particle type, we performed three experiments, each defined by a different (particle) reference buoyancy $b_{z_0}$ [i.e., reference density $\rho_f(z_0)$]. These reference buoyancy values were chosen to match: a value close the bottom of the mixed layer, a value within the thermocline, and a value within the mixed layer. Note that this procedure naturally identifies the vertical coordinate $z_0$ of the isopycnal at which particles are neutrally buoyant. Particles were typically initially distributed randomly in the horizontal, at a specific depth close to their corresponding neutral isopycnals. The above setup allows us to investigate the variability of vertical particle concentration profiles and possible accumulation effects. 

The particle simulations were run for  a dimensionless time $  t = 1.89 \times 10^{4}$. Based on the large-eddy turnover time measured a posteriori in the mixed layer, this corresponds to approximately  $841$ eddy turnover times. The simulations are performed using a Froude number $Fr=1.66 \times 10^{-5}$. Since the inertial acceleration term $D\bm{u}/Dt$ is at most of the same order of the buoyancy term $ b$, the small value of $Fr$ justifies neglecting its contribution in the analysis of particle dynamics in the following sections.

\section{\label{sec3}
Results}
\subsection{\label{subsec31}Eulerian dynamics and mixed layer depth identification}
We begin our analysis by examining the evolution of the stratified flow. The simulation is initialized with a linear buoyancy profile, with a small perturbation added to generate convection \cite{bhamidipati2020turbulent}. Surface cooling establishes negative temperature gradients at the top boundary, causing fluid parcels heavier than their surroundings to sink, while internal heating produces lighter parcels in the interior that tend to rise. This buoyancy-driven overturning circulation leads to the development of descending convective plumes that generate vortices that grow, producing sufficient turbulence to homogenize temperature (and, hence, density) within the upper layer. This  behavior is consistent with field observations of oceanic mixing layers \citep{sprintall2001upper}. The upper, mixed region grows until it reaches a statistically steady state, while below, a thermocline forms, characterized by strong vertical gradients in temperature (and density), and a quite rapid decay of the eddy intensity along the vertical.

\begin{figure}[h!]
\centering
\includegraphics[width=0.65\textwidth]
{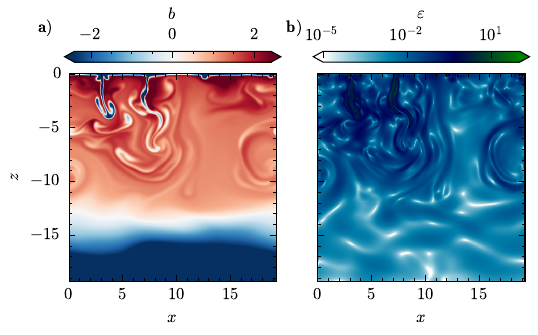}
\caption{Temperature (a) and kinetic energy dissipation rate (b) fields at statistically steady state.}
\label{fig1}
\end{figure}

The mixed-layer depth (MLD) is often defined as the first depth, starting from below, at which vertical temperature gradients approach zero; however, small-scale fluctuations in the simulated temperature field complicate its precise identification in this way, which may lead to an underestimation of the MLD. On the other hand, in our configuration, it might be defined as the first depth from above where the local temperature gradient equals the imposed bottom-boundary gradient; nevertheless, the local temperature gradient evolves during the simulation and no longer remains equal to the imposed value, which can lead to an overestimation of the MLD. For this reason here we adopt an intermediate criterion, identifying the MLD as the first depth where the local temperature gradient reaches $0.7$ times the bottom imposed gradient. At this depth, by inspecting the profiles of the turbulent kinetic energy and temperature gradient, the turbulent kinetic energy has decreased to a low value, while the temperature gradient becomes nearly constant. Figure~\ref{fig1} shows the temperature field and the turbulence dissipation rate once the flow has reached a statistically, essentially steady, state.

\begin{figure}[h!]
\centering
\includegraphics[width=0.95\textwidth]
{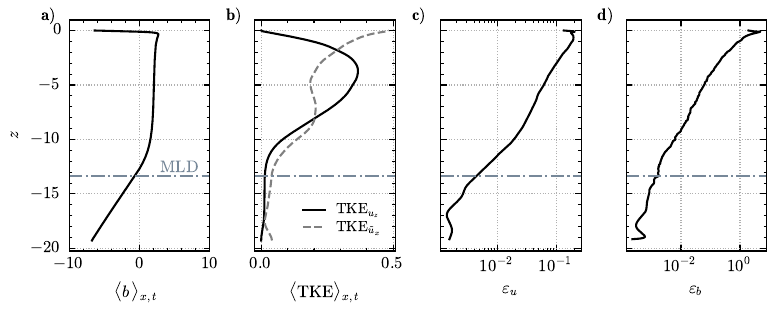}
\caption{Horizontally and temporally averaged buoyancy (a), horizontal and vertical turbulent kinetic energy components (b), turbulence dissipation rate (c), and dissipation rate of buoyancy fluctuations (d) at statistically steady state.}
\label{fig2}
\end{figure}

The horizontally and temporally averaged temperature field initially exhibits a linear profile and evolves over time to reach a steady shape as the one reported in Fig.~\ref{fig2}a. Vertical profiles of the turbulent kinetic energy (Fig.~\ref{fig2}b), turbulence dissipation rate (Fig.~\ref{fig2}c), and buoyancy-variance dissipation rate (Fig.~\ref{fig2}d) reveal the strong vertical inhomogeneity of the flow, with enhanced turbulent activity and scalar mixing within the mixed layer. Both the turbulence and buoyancy-variance dissipation rates become very small below the mixed layer.

\subsection{\label{subsec32}Particle dynamics: confinement and dispersion}
\begin{figure}[h!]
\centering
\includegraphics[width=0.85\textwidth]
{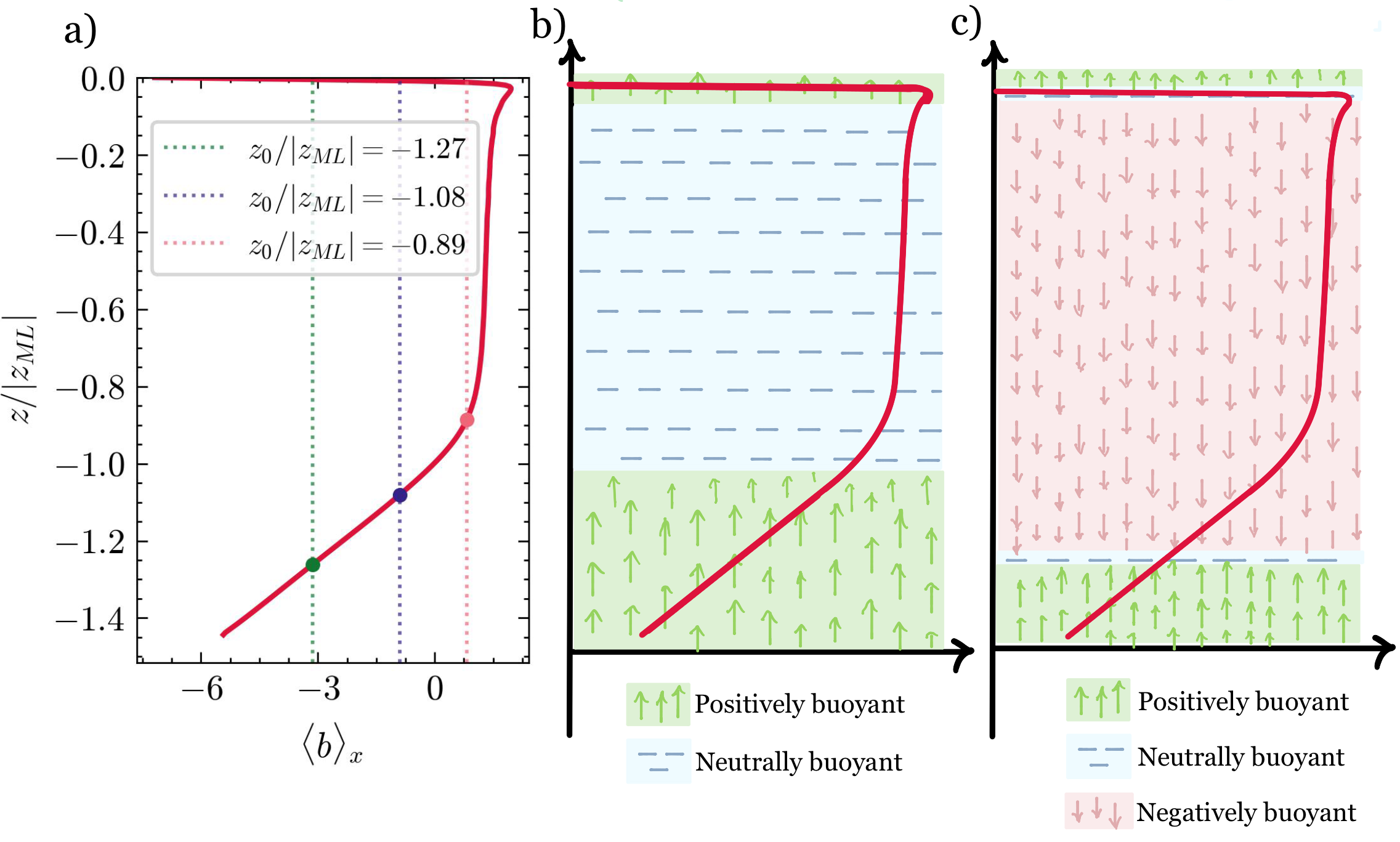}
\caption{
Schematics of the particle simulation settings. (a) Vertical buoyancy profile and the three cases considered, highlighted by the three points, each denoting a different particle reference buoyancy $b_{z_0}$.
Panels (b) and (c) illustrate the regions where particles are positively, neutrally, or negatively buoyant for the cases of particle reference buoyancy values corresponding to a selected depth within the mixed layer (b) or at a specific depth in the thermocline (c).}
\label{fig3}
\end{figure}

Quasi-neutrally buoyant particles, moving with the velocity in Eq.~(\ref{e7}), were released in this flow after it reached statistically steady conditions. Different numerical experiments were considered. Figure~\ref{fig3}a summarizes the corresponding experimental setups, showing the particle reference buoyancy levels and the positions of the isopycnal surfaces $ z_0$ (normalized by the MLD), at which particles are neutrally buoyant, and thus may be expected to accumulate.

Depending on the specific reference buoyancy values, particle behavior changes. Particles with a reference buoyancy equal to that of a selected depth within the mixed layer (see Fig.~\ref{fig3}b) are nearly neutrally buoyant in this region. These particles tend to accumulate near the surface: they rise from the thermocline, are dispersed by turbulent eddies in the mixed layer, and eventually transported to the surface boundary layer (where $\partial_z b<0$), where they become positively buoyant. This surface accumulation effect becomes stronger as the particle response time ($\tau_p$) increases, since the acceleration due to buoyancy scales with particle size [see Eq.~(\ref{e7})]. Smaller particles instead remain in the mixed layer for longer periods, as turbulent transport dominates their dynamics. The situation is different for particles with reference buoyancy equal to that at a specific depth within the thermocline. Two neutrally buoyant depths now exist: one near the surface and another in the thermocline (see Fig.~\ref{fig3}c). However, while the upper one is an unstable fixed point of the Lagrangian dynamics, the lower one is a stable fixed point. Therefore, particles of this type tend to accumulate below the mixed layer. Note that most of the particles that initially remain near the surface are transported by turbulent eddies within the mixed layer, where they eventually become negatively buoyant and sink toward the lower accumulation depth.
\begin{figure}[h!]
\centering
\includegraphics[width=0.95\linewidth]
{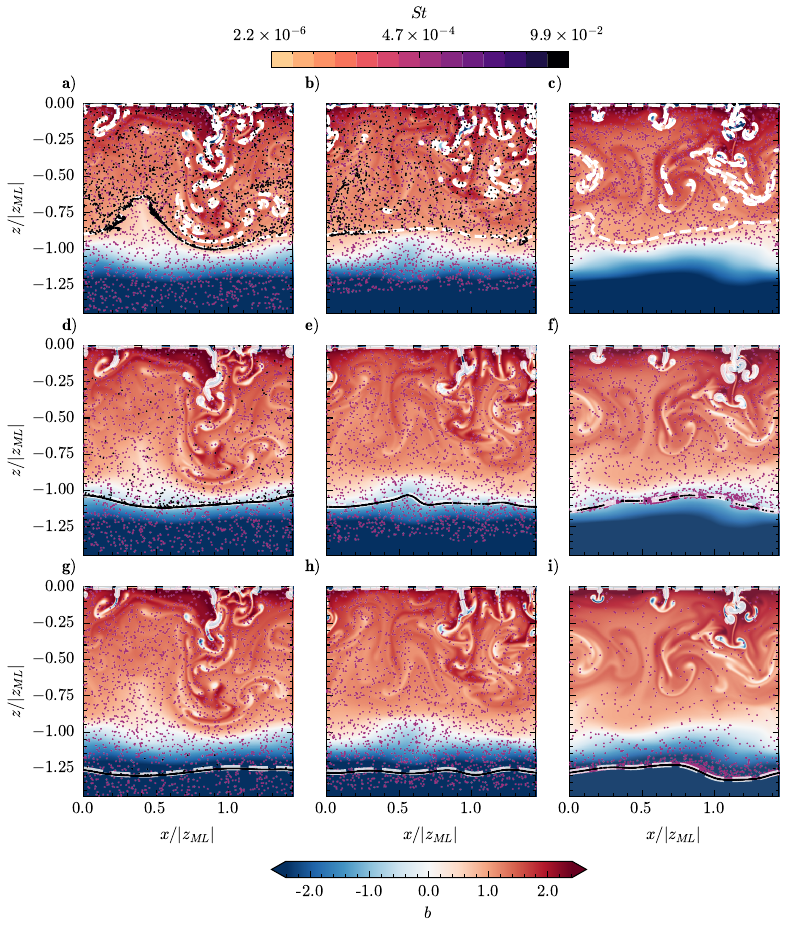}
\caption{Temporal evolution of particle distributions (only the cases of intermediate and largest particle diameters are shown) for different reference buoyancies: buoyancy at a selected depth within the mixed layer (a–c), buoyancy at a depth just below the MLD (d–f), buoyancy at a depth in the thermocline (g–i). White dashed lines indicate the reference isopycnal surfaces. In each row, time grows from left to right; in all cases, particles were seeded at uniformly random initial positions. The simultaneous buoyancy fields are shown in color.}
\label{fig4}
\end{figure}


The transport mechanisms sketched above are confirmed by flow and particle visualizations at different instants of time from our simulations, reported in Fig.~\ref{fig4}. Specifically, panels a, b, and c of Fig.~\ref{fig4} show particle mixing and accumulation, as time increases, near the surface for particles with a reference temperature (or buoyancy) matching the value selected within the mixed layer. Note that the positions of particles represented by dark colors, corresponding to the largest $St$ values, almost perfectly overlap with the upper reference isopycnal (white dashed line), close to the surface.
Panels d to i of the same figure show particle accumulation within the thermocline for particles whose reference buoyancy corresponds to the buoyancy value at the bottom of the mixed layer (d-f) and deeper in the thermocline (g-i), respectively. Also in this case, big enough particles (shown in dark colors) get highly concentrated around the reference isopycnal.

\begin{figure}[h!]
\centering
\includegraphics[width=0.95\linewidth]
{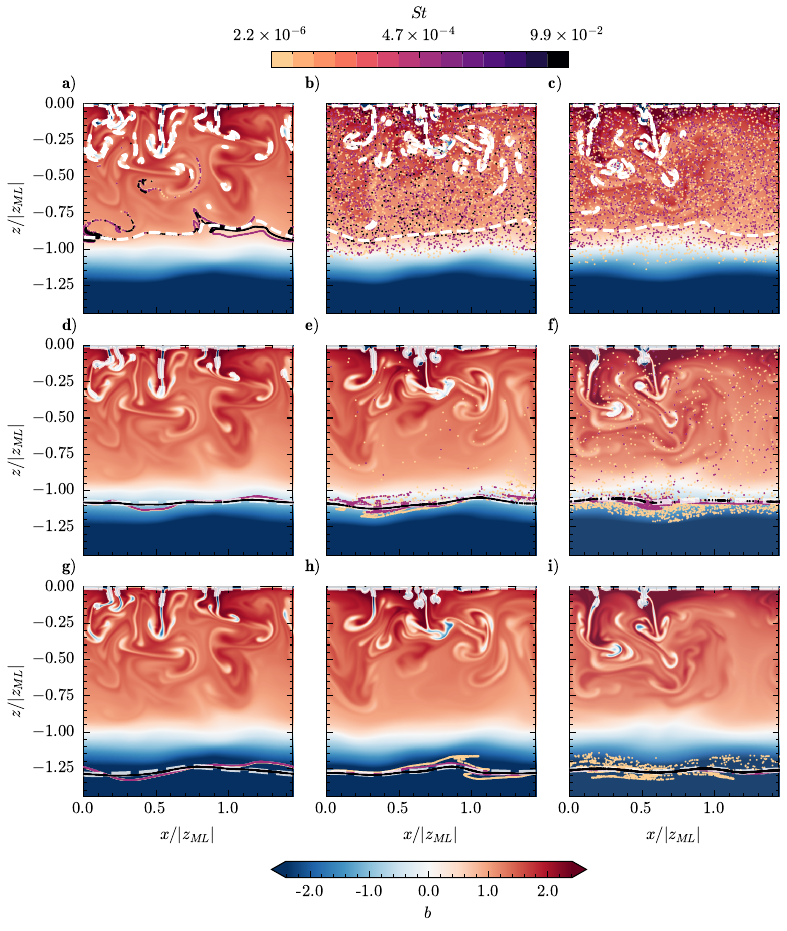}
\caption{Temporal evolution of particle distributions (smallest, intermediate and largest particle diameters) initialized near their accumulation depth for particles within the mixed layer (a–c), close to the mixed layer base (d–f), and within the thermocline (g–i). White dashed lines indicate the reference isopycnals. In each row, time grows from left to right. The simultaneous buoyancy fields are shown in color.
}
\label{fig5}
\end{figure}

To gain more insight into the particle distributions relative to their preferential accumulation depth, we performed further experiments, now releasing particles at random horizontal positions, but with initial vertical positions at their reference depth $ z_0$ (i.e., the one corresponding to $ b_{z_0}$). This configuration allows us to isolate the influence of local flow dynamics at a given depth from the effects associated with an initially vertically uniform particle distribution. As highlighted before, the behavior of particles varies across the different flow regions according to the local turbulence intensity and particle buoyancy response. Particles released within the mixed layer ($| z_0|<| z_{ML}|$, with $ z_{ML}$ the MLD) rapidly adjust to flow motions, become trapped within turbulent eddies, and subsequently accumulate near the surface (Figs.~\ref{fig5}a-\ref{fig5}c). Conversely, particles released 
just below the MLD (Figs.~\ref{fig5}d-\ref{fig5}f) and deeper within the thermocline (Figs.~\ref{fig5}g-\ref{fig5}i) remain attached to their reference isopycnal to varying degrees, depending on their inertial response to turbulence (whose intensity decays vertically below the mixed layer).

\begin{figure}[h!]
\centering
\includegraphics[width=0.95\textwidth]
{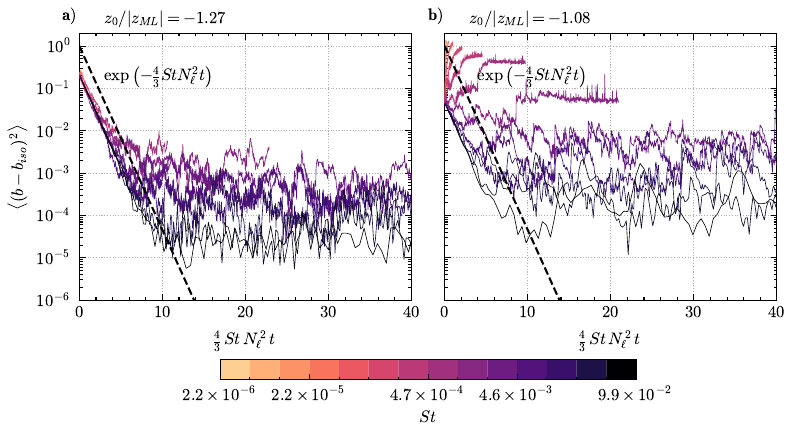}
\caption{Time evolution of the mean squared buoyancy deviation from the reference isopycnal value, $\langle ( b- b_{iso})^2\rangle$ for particles that are neutrally buoyant with respect to a depth in the thermocline (a) and with respect to the MLD (b). Here time is rescaled by $4/3 St N_\ell^2$; see prediction in Eq.~(\ref{e10}).}
\label{fig6}
\end{figure}

Particles initialized in the thermocline and close to the bottom of the mixed layer exhibit more interesting transport dynamics and vertical distribution patterns; therefore, we focus on those cases. In both of them, soon after release, particle trajectories appear different: some particles quickly relax toward their reference isopycnal surface, $ z_{iso}$ (i.e., $\langle  z_{iso}\rangle_{x, t} =  z_0$), while others rapidly develop excursions that gradually move them away from their reference depth. The relaxation towards the reference isopycnal is more evident by looking at the buoyancy experienced by the particles, and can be understood by examining the motion due to the vertical component of the particle velocity $v_z$ 
[from Eq.~(\ref{e8})]. Considering particle motions in the vicinity of the accumulation depth $z_0$ (i.e., $ z_{iso}$), we assume that the fluid velocity depends only on time, so that $u_{z}( t) = dz_{iso}/dt$. Introducing the particle vertical displacement $\zeta =  z - z_{iso}$, performing a Taylor expansion of $b$ around the isopycnal vertical coordinate $z_{iso}$, and neglecting the added mass contribution in the vertical component of Eq.~(\ref{e8}), gives the following evolution for $\zeta$:
\begin{equation}
\frac{d{\zeta}}{d  t} \simeq -\frac{2}{3}St N_{\ell}^2 \zeta,
\label{e9}
\end{equation}
where we set $\partial_{ z}  b|_{ z_{iso}} \simeq \partial_{ z} \langle  b \rangle_{x,t} |_{ z_{iso}}=N_{\ell}^2$, considering that the mean-buoyancy vertical gradient is positive as long as $ z_{iso}$ is below the MLD and can be associated with a  local Brunt-Väisälä frequency. The solution of this equation is $\zeta(t)=\zeta(0)\exp{\left( -2/3 \, St N_{\ell}^2 \,  t \right)}$ and, since $ b- b_{iso} \simeq N_{\ell}^2 \left(  z -  z_{iso} \right)$ [with $ b_{iso}= b( z_{iso})$], one has that
\begin{equation}
\langle \left[  b( z, t)- b_{iso}( z, t) \right]^2 \rangle = \langle \left[  z(0)- z_{iso}(0) \right]^2 \rangle N_{\ell}^4 \exp\left( -\frac{4}{3} St N_{\ell}^2  t \right),
\label{e10}
\end{equation}
where angular brackets denote an average over all particles. According to this reasoning, particle displacements $\zeta(t)$ with respect to the reference isopycnal should vanish after sufficiently long time. As seen in Fig.~\ref{fig6}, to some extent this is confirmed by the measurement of buoyancy at particle positions versus time. Nevertheless, the same numerical results also reveal that if large particles indeed tend to converge toward the isopycnal surface, smaller ones instead, after a while,  disperse around it, as indicated by relatively large values of $\langle \left[b(z,t)-b_{iso}(z,t) \right]^2 \rangle$ at long times.
This effect is more prominent and easier to observe for particles that are neutrally buoyant relative to the bottom of the mixed layer (Fig.~\ref{fig6}b) than for those that are neutrally buoyant with respect to a larger depth in the thermocline (Fig.~\ref{fig6}a). Such dispersion effect suggests that turbulent fluctuations, neglected in the derivation of Eq.~(\ref{e9}) and Eq.~(\ref{e10}), persist below the mixed layer at an intensity that, while reduced, still allows them to play an important role in particle dynamics.

\subsection{\label{subsec33} Particle vertical concentration profiles and their evolution}

\begin{figure}[h!]
\centering
\includegraphics[width=1\textwidth]
{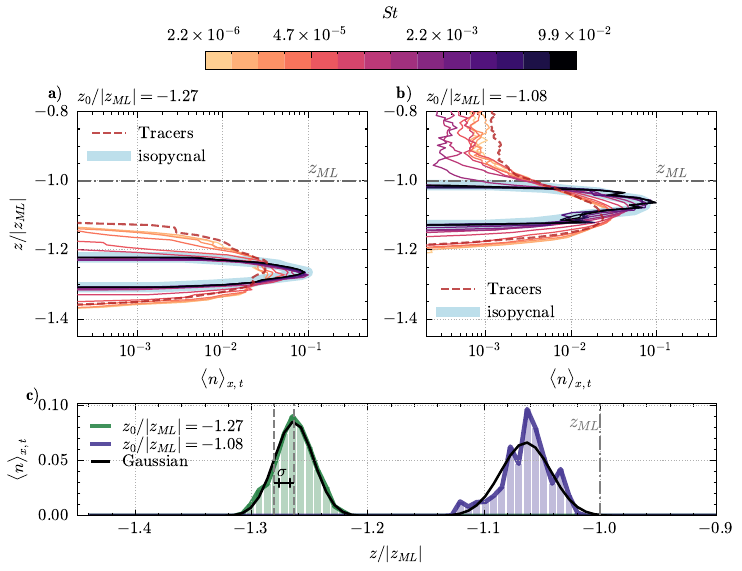}
\caption{Vertical number density profiles for particles neutrally buoyant at depth well inside the thermocline (a) or just below the MLD (b). For both cases, the profiles obtained for the largest particles are compared with Gaussian distributions (having the same mean values and standard deviations as the measured ones) in (c).}
\label{fig7}
\end{figure}

This dispersive behavior and convergence toward isopycnal surfaces can be further analyzed by inspecting the particles’ vertical concentration profiles in the statistically stationary regime, which are shown in Fig.~\ref{fig7}a-b. In general, smaller particles exhibit broader distributions, which are not very different from those of purely Lagrangian tracers ($St=0$), as it appears reasonable, considering their weak inertia [see also Eq.~(\ref{e8})]. Sufficiently large particles, instead, give rise to more pronounced concentration peaks (Fig.~\ref{fig7}a-b). It can be further remarked that the latter distributions are very close to those of the reference-isopycnal depth fluctuations, reflecting particle accumulation along these lines. When the particle reference buoyancy is closer to that at the MLD, dispersion is more important, due to more intense turbulent fluctuations. Small-inertia particles are in this case quite easily entrained into the mixed layer, leading to a strongly asymmetrical distribution characterized by high left tails (Fig.~\ref{fig7}b). For a particle reference buoyancy matching a value at larger depth in the thermocline, the vertical distribution of large particles is not far from a Gaussian, as also observed in previous studies of stably stratified single-layer (i.e. without any mixed layer), externally forced turbulence~\cite{sozza2018inertial}. Moving closer to the mixed layer (in terms of particle reference buoyancy), deviations from Gaussianity emerge, consistent with the presence of more intense turbulence that impacts particle dynamics.

\begin{figure}[h!]
\centering
\includegraphics[width=1.0\textwidth]
{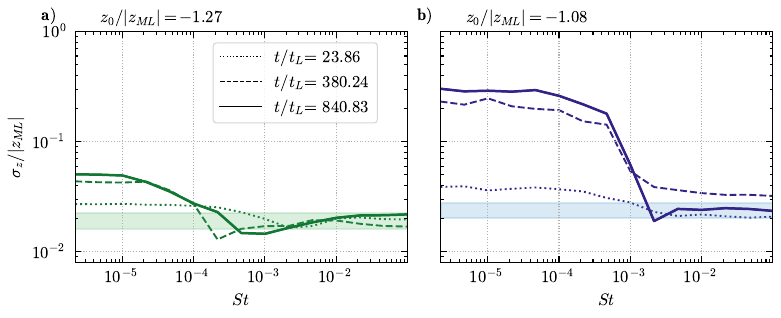}
\caption{Root-mean-square particle displacement from the reference isopycnal $\sigma_z=\langle [ z( t)- z_{0}]^2\rangle^{1/2}$, normalized by the MLD as a function of $St$ at three different times for particles neutrally buoyant at a specific depth in the thermocline (a) and just below the MLD (b). Here, $ t_L$ denotes the large-scale overturning time within the mixed layer. The shaded green and blue regions represent the upper and lower bounds of the isopycnal depth fluctuations over the three instants of time shown.}
\label{fig8}
\end{figure}

An estimate of the typical particle spreading is provided by the width of the concentration profiles, meaning the standard deviation of vertical particle positions, computed with respect to the average preferential accumulation depth $ z_0$, $\sigma_z=\langle [ z( t) -  z_0]^2 \rangle^{1/2}$. The behavior of this quantity as a function of $St$ at several times is reported in Fig.~\ref{fig8} for our two particle-reference-buoyancy cases. As it can be seen, in both cases, for large enough Stokes numbers, $\sigma_z$ is close to the root-mean-square (rms) depth fluctuation of the reference isopycnal, in agreement with the accumulation of large particles around this depth previously observed (Fig.~\ref{fig5} and Fig.~\ref{fig7}). More important spreading (larger $\sigma_z$) is found for smaller particles, which in turn is detected over a broader range of $St$ values, and grows when the particle reference depth is closer to the MLD (compare Fig.~\ref{fig8}a and Fig.~\ref{fig8}b).
This confirms, once more, that the larger turbulence intensity at smaller depths can counteract, to a measurable extent, the buoyancy effects leading to particle accumulation. At the largest times of observation (corresponding to roughly $800$ large-eddy turnover times in the mixed layer), the curves of $\sigma_z(St)$ attain almost steady shapes. Some time dependency seems to be still present, particularly in the case of particle reference buoyancy close to the MLD value (where turbulent motions are more intense), likely due to rare, though possible, events associated with turbulence-driven particle entrainment into the mixed layer, which lead to further growth of $\sigma_z$.


\subsection{\label{subsec34}Predicting particle vertical dispersion through a stochastic model}
To rationalize the above numerical findings, we consider the simplified Lagrangian one-dimensional (1D) model $\dot{ z}=v_{ z}$, with $v_{ z} \simeq u_{ z}-2/3 St ( b- b_{ z_0})$ the vertical component of the particle velocity in Eq.~(\ref{e8}), where we neglected the added-mass term, as the numerics show that this is small. Aiming to inspect the interplay between relaxation to the particle reference isopycnal and turbulent motions, we next decompose the fluid velocity and buoyancy into mean and fluctuating quantities, $u_{ z}= \overline{u_{ z}}  + u_{ z}'$ and $ b = \langle  b \rangle_{x, t} +  b’$. Here, the overbar indicates a temporal average over Lagrangian trajectories and $\langle \dots \rangle_{x, t}$ an average of the Eulerian field over the horizontal coordinate and time; if the averaging time is sufficiently long, $\overline{u_{ z}} \approx \langle  u_{ z} \rangle_{x, t} \approx 0$. Within this framework, the particle vertical motion is then described by
\begin{equation}
\frac{d z}{d t} = u_{ z}’ - \frac{2}{3} St \left( \langle  b \rangle_{x, t} +  b’ -  b_{ z_0}\right).
\label{e11}
\end{equation}
Since we are interested in the dynamics for particles close to the preferential accumulation depth $ z_0$, we develop the mean buoyancy field to first order as
$\langle  b \rangle_{x, t} \approx \langle  b\rangle_{x, t}( z_0) + N_{\ell}^2( z- z_0) \approx  b_{ z_0} + N_{\ell}^2( z- z_0)$ (recall that $N_{\ell}^2=\partial_{ z} \langle  b \rangle_{x,t} |_{ z_{iso}} \simeq \partial_{ z} \langle  b \rangle_{x,t} |_{ z_0}$, and note that the buoyancy profile is essentially steady, and nearly linear in the thermocline). Then, if one further calls $\xi(t)= z(t)- z_0$ the particle displacement, to first order in $\xi$, Eq.~(\ref{e11}) can be rewritten as
\begin{equation}
    \frac{d\xi}{d t} = -\mu \xi + u_{ z}' - \frac{2}{3} \, St \,  b',
\label{e12}
\end{equation}
where $\mu= 2/3 \, St \, N_{\ell}^2$. From Eq.~(\ref{e12}) it becomes evident that particles are subject to a deterministic drift  (induced by the mean buoyancy profile) that recalls them to their reference depth $ z_0$ and (fluid-velocity and buoyancy) random fluctuations that tend to disperse them around it.
Note that the intensities of both effects depend on particle inertia ($St$).

The rationale behind the above modeling approach shares some similarities with that of~\cite{sozza2018inertial}, where a stochastic model in particle position-velocity space was proposed to investigate the analogous problem of particle spreading in a single, stably stratified fluid layer. Equation~(\ref{e12}) can then be seen as a reduced version (i.e., in position space only) of the model from that study. However, it is worth remarking that in a stably stratified fluid, fluctuations in the form of both turbulent motions and internal gravity waves~\cite{Sutherland_2010} coexist. The role of the latter was not examined in~\cite{sozza2018inertial}, possibly due to the forced nature of the turbulent flows considered there, which may mask the contribution of internal waves. In the present case, the similarity of the particle and isopycnal-depth-fluctuations vertical distributions (see Fig.~\ref{fig7}) suggests that internal gravity waves should also be taken into account. Therefore, we now further decompose fluctuations as $u_{ z}’ = u_{ z_w}’ + u_{ z_n}'$ and $ b’ =  b_{w}’ +  b_{n}'$, where subscript $w$ indicates a more coherent, wavy component and subscript $n$ a more noisy, turbulent one.

The form of the perturbations due to internal gravity waves can be obtained from the  non-dissipative ($\nu=\kappa=0$) and unforced ($Q_0=0$), linearized Boussinesq equations (see, e.g.,~\cite{Sutherland_2010} for more details).
Restricting the attention to vertical dynamics only, and neglecting pressure in the vertical momentum equation, as it can be expected to scale as the nonlinear term, one has:
\begin{subequations}
\begin{align}
\frac{\partial u_{ z_w}'}{\partial  t} &=  b_w'\\
\frac{\partial  b_w'}{\partial  t} &= -N_{\ell}^2 u_{ z_w}' \, .
\end{align}
\label{e13}
\end{subequations}
Taking the time derivative of Eq.~(\ref{e13}a) and plugging it into Eq.~(\ref{e13}b) gives $\partial_{ t}^2 u_{ z_w}'+ u_{ z_w}'N_{\ell}^2 =0 $, from which the expressions of the velocity and buoyancy fluctuations associated with the vertical motion of the reference isopycnal can be obtained:
\begin{subequations}
\begin{align}
u_{ z_w}' &= A_w \sin{(N_{\ell} t+\phi)},\\
 b'_w&=A_w N_{\ell}\cos{(N_{\ell} t+\phi)}.
\end{align}
\label{e14}
\end{subequations}
Here $A_w$ and $\phi$ denote the amplitude and the phase (which depends on the particle initial position) of the wave-induced vertical velocity fluctuations, respectively, which can be obtained from our data. Note that in the above derivation, for simplicity, we are assuming that $\bm{k}_w \cdot \bm{x}=\mathrm{const}$, with $\bm{k}_w$ the wave vector, coherent with our choice to describe small particle displacements from $ z_0$.

As for turbulent fluctuations, we model them through Gaussian white noise terms ($\eta_u, \eta_b$). Their amplitudes can be determined according to the values of the eddy-diffusion coefficients associated with buoyancy ($D_b$) and advection ($D_u$) effects, measured from the numerical data
(see~\cite{chavez2026turbulent} for a recent discussion on the possible methods to do so, and their limitations). In other words, for the noisy (meaning turbulent) terms we take $u_{z_n}’ = \sqrt{2D_u}\eta_u(t)$, $b_{n}’ = \sqrt{2D_b}\eta_b(t)$. Using the above approximations, Eq.~(\ref{e12}) for particle displacements then becomes:
\begin{equation}
\frac{d\xi}{d t} = -\mu \xi + A_w \sin{(N_{\ell} t+\phi)} - A_w \frac{\mu}{\tilde{N_{\ell}}}\cos{(N_{\ell} t+\phi)}  + \sqrt{2D_u}\eta_u(t) - \frac{\mu}{N_{\ell}^2}\sqrt{2D_b}\eta_b( t).
\label{e15}
\end{equation}

This is the evolution equation of a stochastic process with deterministic drift $-\mu \xi$ and noise $\sqrt{2D_u}\eta_u(t) - \mu/N_{\ell}^2\sqrt{2D_b}\eta_b(t)$, i.e., an Ornstein-Uhlenbeck process~\cite{gardiner2004handbook}, additionally forced by the periodic terms originating from internal gravity waves.
The formal solution of Eq.~(\ref{e15}) is
\begin{equation}
\xi (t)= \xi_0 \, e^{-\mu  t} + \frac{A_w}{N_{\ell}}e^{-\mu  t} \cos{\phi} -
\frac{A_w}{N_{\ell}}\cos{\left(N_{\ell} t+ \phi\right)} + \int_0^{ t} e^{\mu(s- t)}
\left[ \sqrt{2D_u}\eta_u(s)  -\frac{\mu}{N_{\ell}^2} \sqrt{2D_b}\eta_b(s) \right] \, ds
\end{equation}
[with $\xi_0=\xi(0)$], from which the variance of the particle displacement from the reference depth $ z_0$ can be computed. Considering that particles are initially uniformly randomly distributed along the horizontal at depth $ z_0$ (i.e. $\xi_0=0$ for all of them), and that averaging over the ensemble of their initial conditions ($\langle \dots \rangle_e$) is then equivalent to averaging over the phase $\phi$ (in the terms originating from wavy motions), one obtains
\begin{equation}
\langle \xi^2 \rangle_e = \frac{A_w^2}{2 N_{\ell}^2}\left[ 1+e^{-2\mu t}-2 \, e^{-\mu t}\cos{(N_{\ell} t)}\right] +
\left( \frac{D_u}{\mu}+ \frac{\mu D_b}{N_{\ell}^4}\right)\left( 1-e^{-2\mu t}\right).
\label{e16}
\end{equation}
Here we further used the fact that the cross-term involving the product of velocity and buoyancy fluctuations is negligible, as suggested by numerical simulations. More details can be found in App.~\ref{AppA}.

The above expression reveals the transient dynamics of the particle displacement from the preferential accumulation depth $ z_0$. In particular, it highlights the two competing mechanisms governing vertical particle transport over time: on one side, $\langle \xi^2 \rangle_e$ relaxes toward the variance associated with isopycnal oscillations, $A_w^2/(2N_{\ell}^2)$; on the other, the effective diffusion induced by turbulent buoyancy and velocity fluctuations becomes more relevant as time increases. We can further remark that, excluding the contribution from waves, and in the long time limit, Eq.~(\ref{e16}) describes a non-monotonic behavior in $St$, $3D_u/(2 N_\ell^2 St) + 2(D_b/N_\ell^2) St$ (recall that $\mu=2/3 St N_\ell^2$), akin to the one reported in Ref.~\cite{sozza2018inertial}. However, it should be noted that the particle diameters considered in the present study are always quite small, so that $St<0.1$. Therefore, it appears reasonable to neglect the contribution from $D_b$ with respect to that from $D_u$.

To compare the prediction in Eq.~(\ref{e16}) with the numerical data, we now need to estimate $A_w$ and $D_u$. The contribution of wave-induced vertical velocity fluctuations can be estimated directly from the variance of the isopycnal displacement, which can be obtained by solving $\dot{ z}_{iso}=u_{ z_w}'$ with $u_{ z_w}’$ from Eq.~(\ref{e14}a),
\begin{equation}
\langle \xi_{\rm iso}^2\rangle_e = \frac{A_w^2}{N_{\ell}^2} \left[1-\cos(N_{\ell} t)\right].
\label{eq:xi_iso_squared}
\end{equation}
Further averaging this expression over the wave period $T_w = 2\pi/N_{\ell}$ gives
\begin{equation}
A_w^2 = \langle \xi_{\rm iso}^2\rangle_{e,T_w} N_{\ell}^2,
\label{eq:A_w_suqared}
\end{equation}

where the subscript $T_w$ indicates the additional time average.

\begin{figure}[h!]
\centering
\includegraphics[width=1\textwidth]
{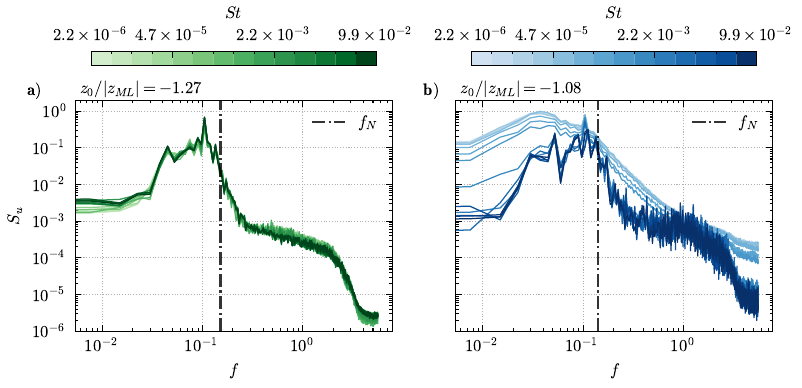}
\caption{Temporal spectra of particle velocity fluctuations for particles released in the thermocline (a) and near the base of the mixed layer (b). The vertical dashed line indicates the local buoyancy frequency, $f_N = N_{\ell}/(2\pi)$.
}
\label{fig9}
\end{figure}

On the other hand, estimating particle diffusion coefficients is more challenging because fluctuations along Lagrangian trajectories contain contributions from both turbulence and internal gravity waves, which may overlap over certain frequency ranges. Therefore, an exact separation between these two contributions is generally not possible. Nevertheless, most of the energy associated with internal gravity waves is concentrated at frequencies smaller than the buoyancy frequency~\cite{d2000lagrangian, d2000wave}. The signature of waves can then be appreciated by inspection of the temporal spectra of particle velocity fluctuations $S_{u}$ (shown in Fig.~\ref{fig9}). For particles with reference-buoyancy depth well inside the thermocline (Fig.~\ref{fig9}a), the spectrum is dominated by a peak at frequencies $f<N_\ell/(2\pi)$ for all $St$ numbers, confirming that in this case waves play a major role in the Lagrangian dynamics, in agreement with the already observed important localization of particles (of all types) close to the reference isopycnal. When the particle reference-buoyancy depth is close to the base of the mixed layer (Fig.~\ref{fig9}b), instead, this is true only for large enough particles. For smaller particles, which can be more easily entrained into the mixed layer, the spectral peak becomes  progressively less well defined as $St$ decreases, and a broadband spectrum, typical of turbulent dynamics, develops. These observations motivate estimating the particle diffusion coefficient according to the flow region sampled by the particles along their trajectories, particularly for those released near the bottom of the mixed layer.

Inertial particle dynamics reduce to tracer dynamics in the limit of small particle diameter (i.e., for $\mu=0$). Therefore, a possibility to estimate the diffusion coefficient due to $u_{ z_n}'$ can be offered by the examination of the mean square displacement of tracers initially released at the same positions as those of the inertial particles. Performing a Taylor expansion around $\mu=0$ of the exponential terms in Eq.~(\ref{e16}) allows to obtain the tracer displacement variance $\langle \xi_{\rm tracer}^2 \rangle_e = \langle \xi_{\rm iso}^2\rangle_e + 2D_u  t$, from which a time-dependent diffusion coefficient can be computed as:
\begin{equation}
D_u( t) = \frac{\langle \xi_{\rm tracer}^2 \rangle_e -\langle \xi_{\rm iso}^2\rangle_e }{ 2  t} , .
\label{eq:Du_time}
\end{equation}

\begin{figure}[h!]
\centering
\includegraphics[width=1\textwidth]
{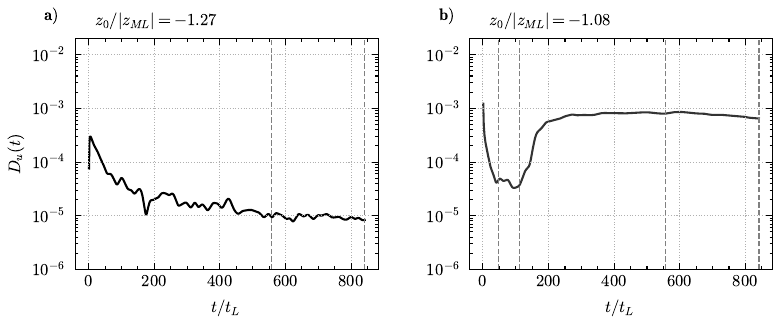}
\caption{Time-dependent diffusion coefficient $D_u( t)$ [see Eq.~(\ref{eq:Du_time})] for tracers released in the thermocline (a) and just below the mixed layer (b). In each panel, the vertical dashed lines indicate the averaging intervals used to estimate $D_u$. Below the thermocline, a single diffusion coefficient is obtained, whereas for tracers released at the bottom of the mixed layer, two coefficients are estimated according to the region of the flow sampled by the particles.
}
\label{fig10}
\end{figure}

Once this quantity approaches a constant value, $D_u$ will be obtained from the temporal average of $D_u( t)$ over the time interval corresponding to such plateau region. The results are presented in Fig.~(\ref{fig10}). Below the thermocline, a single diffusion coefficient can be quite safely identified at late times (Fig.~\ref{fig10}a). For tracers released near the bottom of the mixed layer, we obtain two possible estimates of the diffusion coefficient, which we associate with particles that remain in the thermocline (lower value at shorter times), and with particles that escape to the mixed layer (larger value at longer times), respectively (Fig.~\ref{fig10}b), consistent with the different flow regimes detected in the temporal spectra of Fig.~\ref{fig9}.

\begin{figure}[h!]
\centering
\includegraphics[width=0.90\textwidth]
{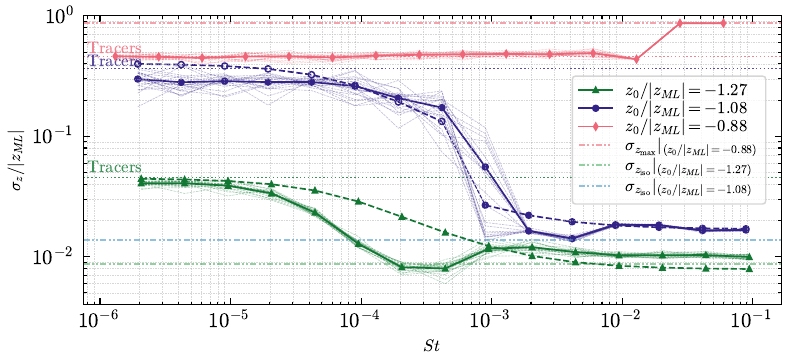}
\caption{Typical particle displacement $\sigma_z$ measured from the numerical data (solid lines and points), and model prediction from Eq.~(\ref{e16}) (dashed lines and points) for reference buoyancy depth in the thermocline ($ z_0/| z_{ML}| = -1.27$) and
just below the MLD ($ z_0/| z_{ML}| = -1.08$).
The case of reference buoyancy corresponding to the mean value in the mixed layer ($ z_0/| z_{ML}| = -0.88$) is also shown for comparison. All curves are shown at time $ t/ t_L = 836.5$, with $ t_L$ the large-scale eddy-turnover time in the mixed layer. Note that for the case $ z_0/| z_{ML}| = -1.08$, empty and filled circle markers correspond to the prediction computed using the diffusion coefficient, $D_u$ obtained from a time average in the intervals $47 <  t/ t_L < 111$ and $556 <  t/ t_L < 841$, respectively. For the numerical data, the lighter dashed lines show $\sigma_z/| z_{ML}|$ computed independently for subsets of $100$ particles, illustrating the variability of the ensemble averaged particle displacement.
Dotted horizontal lines indicate the value of $\sigma_z$ measured from tracers initialized as inertial particles, in each of the three cases (corresponding to the different values of $ z_0$).}
\label{fig11}
\end{figure}

Having the amplitude of the wave-induced vertical velocity fluctuations $A_w$ and the turbulent diffusivity $D_u$, it is now possible to compare the prediction in Eq.~(\ref{e16}) with the measurement of the particle displacement variance from the numerical data. The results for the numerical and theoretical estimates of $\sigma_z = \sqrt{\langle \xi^2 \rangle_e}$ (normalized by $\vert  z_{ML} \vert$) are shown in Fig.~\ref{fig11} at a given, large time. We further checked that these are robust with respect to the choice of the instant of time. The prediction from the simplified 1D model captures the qualitative behavior of the particle displacement as a function of $St$ for both the reference buoyancy depths, $ z_0/ |z_{ML}|=-1.27$ and $ z_0/ |z_{ML}|=-1.08$. The agreement is quantitatively quite good, particularly for the largest and smallest Stokes numbers, for which particle motions are dominated by waves and turbulence, respectively. Incidentally, these observations also confirm, a posteriori, that the choice to neglect $D_b$ in Eq.~(\ref{e16}) is justified. Some deviations are nevertheless visible, especially for intermediate $St$ values, which may be ascribed to the stronger interplay between the different transport mechanisms involved, in this range of particle diameters, and to the different hypotheses made in the construction of the model.

Among the latter, we argue that the monochromatic nature of waves (only partially supported by the broad peak of the spectrum in Fig.~\ref{fig9}, rather pointing to a wider frequency distribution), as well as the neglect of the horizontal wave structure, likely play a role. At the same time, the assumption of a constant diffusivity, in spite of the evident vertical dependence of turbulent intensity, cannot be considered always fully appropriate (particularly for particles that are close to the mixed layer and can escape into it) and should be rather taken just as a working hypothesis. Within these caveats, the model reveals however, quite successful in accounting for the typical spreading of particles around their reference isopycnal.

To complete the picture, in Fig.~\ref{fig11} we also show $\sigma_z$ for particles with reference buoyancy corresponding to the selected depth within the mixed layer ($ z_0/| z_{ML}|=-0.88$) and purely Lagrangian tracers. In the first case, due to the large turbulence intensity in the mixed layer, particles are quite well mixed (relatively large value of $\sigma_z$) and quite insensitive to buoyancy recalling effects, except at the largest values of $St$. For the largest particle diameters, however, as expected (recall the discussion of the case in Fig.~\ref{fig3}b), they start to accumulate at the surface, which leads to a maximum value of $\sigma_z/\vert  z_{ML}\vert \simeq 1$. As for tracers, released with the same initial conditions as inertial particles, the typical displacement $\sigma_z$ (horizontal dotted lines in Fig.~\ref{fig9}) is close to that of the smallest particles, for each of the three $ z_0$ cases.
Since this value is considerably smaller than the one expected under completely homogenized conditions ($h^2/\sqrt{12} \simeq 107$), which tracers should attain after sufficiently long time, this fact highlights that the system is still far from its long-time, fully mixed state. This also means that, while our results for inertial particles appear essentially time-independent, their extrapolation to the infinite-time limit should be treated with caution.

We end this section by commenting on the vertical dependence of the effective diffusion coefficient, and its implications for a full interpretation of the numerical results using the above modeling framework in the very long time limit. Indeed, due to the variation of turbulent intensity with $z$, along their trajectories, particles of the same type may sample flow regions with 
different diffusivity (in particular, considerably larger inside the mixed layer). However, our simplified 1D model does not account for this feature, as it employs a constant diffusion coefficient, which limits the possibility to investigate the asymptotic state, in time, of the system.
Characterizing the latter numerically can also be prohibitive, because extremely long simulations are needed, due to the importance of quite rare events, as we shall discuss below.

\begin{figure}[h!]
\centering
\includegraphics[width=0.9\textwidth]
{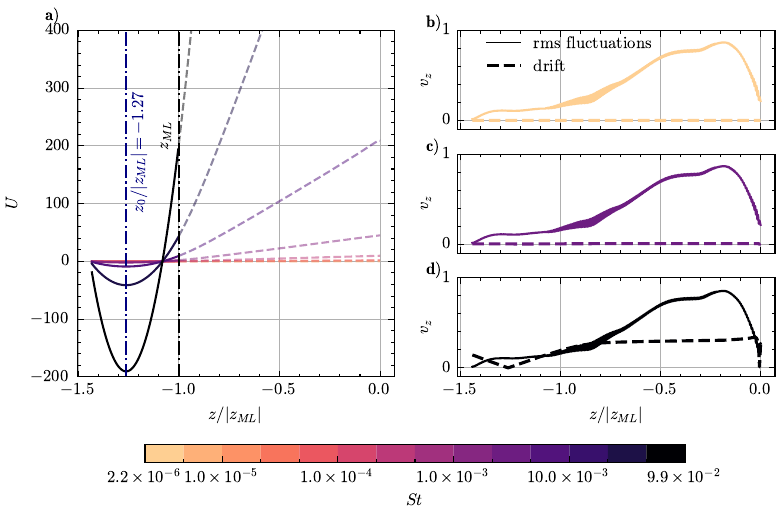}
\caption{(a) Potential $U$ as a function of $z/ |z_{ML}|$ for different Stokes numbers (see color coding). Comparison between the rms turbulent fluctuations and the deterministic drift (see text for definitions), as a function of the vertical coordinate, for small (b), intermediate (c), and large (d) particle diameters corresponding to $St = \num{2.28e-6}$, $\num{2.15e-3}$, and $\num{9.94e-2}$, respectively.}
\label{fig12}
\end{figure}

To illustrate the situation, it is useful to introduce the potential $U( z)=-\int a( z) d z$, associated to the drift term $a( z)=-2/3 \, S_t \, (\langle  b \rangle_{x, t} -  b_{ z_0})$ in Eq.~(\ref{e11}).   This is shown in Fig.~\ref{fig12}a; note that its shape is almost parabolic [see also Eq.~(\ref{e12})]. Accumulation around the reference isopycnal at $ z_0$ is then equivalent to particles being trapped in the potential well (with minimum in $ z_0$). While particles typically wander randomly around this depth due to turbulent fluctuations
[$u_z’-2/3 \, St \,  b’$ in Eq.~(\ref{e11})], a sufficiently intense fluctuation could let them reach the mixed layer, where in turn turbulent motions dominate. Clearly, this is easier when $ z_0$ is closer to the MLD, because in this case the potential barrier to overcome is shallower. However, it is in principle possible, and indeed observed in the simulations, also when the reference isopycnal is at a larger depth in the thermocline, and we will now focus on this case for reference. Note that the time for this to occur can be extremely long; indeed, it may be expected to grow exponentially with the ratio $\Delta U/D_u$~\cite{gardiner2004handbook}, where $\Delta U=U(z_{ML})-U( z_0)$.
Independently of this, once a particle reaches the mixed layer, its fate depends on the relative importance of the drift and noise terms there. Diffusivity is much larger in the mixed layer and, numerically, we find that it is also definitely larger than the drift for all $St$ values (Fig.~\ref{fig12}b). This means that the recalling, deterministic term, while present, is very weak and becomes essentially ineffective. Consequently, after they enter the mixed layer, particles spend a very long time inside it. The probability to escape from $ z_0$ to the mixed layer grows with decreasing particle diameter, because the potential well is shallower (as $St$ is smaller), so that less intense fluctuations suffice to exit the well. As shown in Figs.~\ref{fig12}b and c, for sufficiently small particles, the rms fluctuations are larger than the drift also below the mixed layer, indicating a higher likelihood of particle escape, and hence the possibility of a partial emptying of the accumulation depth, in these cases.

The reasoning sketched above suggests that the mixed layer might be seen as an absorbing barrier for particle dynamics, with larger consequences on smaller particles. Nevertheless, another feature further complicates the picture. The larger diffusivity in the mixed layer, with respect to that of the thermocline,
might also trigger some reverse transitions. In other terms, some large fluctuation in the mixed layer might push a particle back to some depth in the thermocline, where the drift becomes effective again to retain the particle close to the reference isopycnal. To gain full insight into the particle distribution at very large times, both this effect and the previous one should be taken into account, which is far from trivial.

Summarizing, while the present approach allows to identify the key mechanisms controlling accumulation and spreading around the reference isopycnal, particle dynamics in the long-time limit seem to be highly complex and remain to be fully characterized.
A possibility to enhance their understanding may be offered by even more idealized, Langevin-type models, constructed in analogy with Eq.~(\ref{e15}), where one could systematically explore the role of the different parameters, taking into account their spatial variability. A careful investigation of this point would require a dedicated study and extensive simulations, and is thus left for future work.

\section{\label{sec4}Conclusions}

In this work, we investigated, by means of DNS, the vertical dispersion of weakly inertial, quasi-neutrally buoyant particles, whose density matches the fluid one at a given depth, in an idealized ocean convective flow. The flow is characterized by a pronounced nonhomogeneity in the vertical, due to the presence of an upper, essentially homogeneous, mixed layer and a deeper, strongly stratified, thermocline. We showed that particles tend to accumulate around the depth corresponding to their density, and that their vertical displacements and concentration profiles strongly depend on the position of such depth with respect to the two-layer structure of the flow. 

In the thermocline, where stratification dominates and vertical motions are damped, our results agree with previous work on single-layer stably stratified turbulence, highlighting that vertical transport is strongly suppressed. Moreover, depending on particle inertia, this suppression gives rise either to bounded vertical dispersion associated with internal gravity waves~\cite{nicolleau2000turbulent, magnier2026lagrangian}, or to the accumulation and organization of particles into thin layers around their reference isopycnal surfaces~\cite{sozza2016large,sozza2018inertial}. This confinement is controlled by the mean component of buoyancy effects, which acts as a restoring force. Internal gravity waves induce vertical, periodic, quasi-monochromatic displacements that modulate the structure of the vertical concentration profile, while weak turbulent velocity fluctuations account for particle spreading around the equilibrium depths. As the particle response time decreases (hence for smaller sizes), particles become more sensitive to turbulent fluctuations, which results in broader concentration profiles around the reference, equilibrium depth. 

When the preferential accumulation depth gets closer to the mixed layer (approaching it from below), concentration profiles broaden relative to those deeper in the thermocline, and high tails emerge on their small-depth side, reflecting the possible escape of particles into the mixed layer, triggered by turbulent velocity fluctuations. In the special case in which the reference buoyancy corresponds to a selected depth within the mixed layer, particles are transported by turbulent eddies and eventually get trapped at the surface, since they become positively buoyant in the thin boundary layer at the top of the system. 

The numerical findings on the typical spreading of particles around their reference isopycnal are understood by introducing a 1D stochastic model. The latter is obtained from the basic particle equations of motion after a decomposition into mean and fluctuating velocity components, with the first accounting for the buoyancy-driven attraction to the reference depth, and the latter for internal-wave and turbulent motions. For the sake of simplicity, in this model we neglect the vertical dependence of the turbulent diffusivity, and we only consider monochromatic waves, which allows obtaining a relatively simple prediction for the rms particle displacement. In spite of these simplifications, the model provides a satisfactory description of the numerical data as a function of Stokes number, both for reference depths well inside the thermocline and close to the mixed layer.

We note, however, that the assumption of a vertically independent turbulent diffusivity poses some limitations on the possibility of studying the asymptotic time limit of the system, particularly at small Stokes numbers. Indeed, while our simulations cover durations of several tens to hundreds of large eddy turnover times, addressing rare events associated with the variation of the diffusivity experienced by particles along their trajectories, which can induce transitions from one layer to the other, remains challenging. An interesting perspective to address this point could be to consider a stochastic two-layer toy model, which is left for future work.

Overall, the proposed model successfully captures the main physical mechanisms governing particle transport below the mixed layer and highlights how the competition between stratification via wave-induced trapping and residual turbulence controls the vertical spreading of quasi-neutrally buoyant particles in this idealized, but vertically inhomogeneous, convective ocean flow. This work then provides a better understanding of the transport of inertial particles in the intermediate Stokes-number regime in nonhomogeneous turbulence, which may be relevant for the correct sampling of marine plastic pollution and for better accounting for the proportion of plastics across the different regions of the water column, thereby contributing to help closing the mass budget of plastics in the sea~\cite{poulain2024laboratory}. In addition, such improved  understanding of the interplay between different physical processes controlling the accumulation of particulate material may also be of interest for the study of thin phytoplankton layers, complementing the information on the biological dynamics of motile species. Such layers play a key role in marine ecosystems by sustaining organisms belonging to higher trophic levels~\cite{durham2012thin}.  

In perspective, it could be interesting to further enhance the oceanographic realism of our simulations. One viable approach would be to include the wind forcing at the surface, which increases turbulence and shear, as well as turbulent diffusion, and can then modify the relative importance of turbulent and wave-driven transport. Remarkably, this would also allow exploring situations typical of different seasons, since winter situations are generally characterized by more intense mechanical forcing at the surface than summer ones~\cite{callies2016role,berti2021lagrangian}.
Another interesting extension would be to investigate particles lighter or heavier than the surrounding fluid in the present stratified-flow configuration. This may allow accounting for the behavior of a broader class of materials present in the water column.

\appendix

\section{\label{AppA} 
Particle displacement variance in the 1D stochastic model}
In this appendix, we provide more details about the derivation of the expression for the evolution of the particle displacement variance $\langle \xi^2\rangle_e$  
in Eq.~(\ref{e16}) First, recall that the solution of Eq.~(\ref{e15}) is 
\begin{equation}
\xi(t) = \xi_0e^{-\mu  t} + 
\frac{A_w}{N_{\ell}}e^{-\mu  t} \cos{\phi} -
\frac{A_w}{N_{\ell}}\cos{\left(N_{\ell} t + \phi \right)} + \int_0^{ t} e^{\mu(s-t)} \left[ u_{ z_n}'(s) -\frac{\mu}{N_{\ell}^2} b_n'(s)\right] \, ds \, ,
\label{eq:sde_sol_app}
\end{equation}
where 
$u_{z_n}' = \sqrt{2D_u}\eta_u(t)$ and $b_{n}' = \sqrt{2D_b}\eta_b(t)$ (see Sec.~\ref{sec3}). For simplicity, 
we rename the different terms in Eq.~(\ref{eq:sde_sol_app}) as follows:
\begin{align*}
    \xi_{tr}( t) & = \xi_0e^{-\mu  t} + \frac{A_w}{N_{\ell}}e^{-\mu  t} \cos{\phi} \, , \\
    \xi_w( t)  & = -
  \frac{A_w}{N_{\ell}}\cos{(N_{\ell} t+ \phi)} \, ,\\
  \xi_n( t) & = \int_0^{ t} e^{\mu(s- t)} \left[u_{z_n}'(s) -\frac{\mu}{N_{\ell}^2} b_n'(s) \right] \, ds \, .
\end{align*}

The variance of the particle displacement $\xi$ is, clearly, 
\begin{equation*}
\langle \xi^2\rangle_e = \langle \xi_{tr}^2\rangle_e + \langle \xi_{w}^2 \rangle_e + \langle \xi_{n}^2\rangle_e + 2\left( \langle \xi_{tr}\xi_{w}\rangle_e + \langle \xi_{tr}\xi_{n}\rangle_e + \langle \xi_{w}\xi_{n}\rangle_e \right) \, ,
\end{equation*}
where averaging over the ensemble of particles is the same as  averaging over 
the phase $\phi$ for the wavy terms, as particles are uniformly distributed at random horizontal positions, and all at a depth equal to $ z_0$ (which, of course, also implies $\xi_0=0$), at time $ t=0$. In the above expression, among the cross-terms, those involving noise ($\xi_n$) average to zero, so that the only remaining one is
\begin{equation}
    \langle \xi_{tr}\xi_{w}\rangle_e= -\frac{A_w^2}{2N_{\ell}^2}e^{-\mu t}\cos{\left(N_{\ell} t\right)}.   
    \label{eq:cross-term}
\end{equation} 
The first two deterministic contributions (with $\xi_0=0$) amount to
\begin{equation}
    \langle \xi_{tr}^2 \rangle_e = \frac{A_w^2}{2N_{\ell}^2} e^{-2\mu t} \, , \quad    
    \langle \xi_{w}^2 \rangle_e = \frac{A_w^2}{2N_{\ell}^2} \, .
    \label{eq:var_trans_wave}
\end{equation}
The remaining stochastic contribution is
\begin{equation}
\begin{split}
\langle \xi_n^2\rangle_e = \int_0^{ t}\int_0^{ t}e^{\mu(s_1+s_2-2 t)} \langle u_{ z_n}'(s_1)u_{ z_n}'(s_2)\rangle_e \, ds_1ds_2 \\
 + \frac{\mu^2}{N_{\ell}^4}\int_0^{ t} \int_0^{ t}e^{\mu(s_1+s_2-2 t)} \langle  b_n'(s_1)  b_n'(s_2)\rangle_e \, ds_1ds_2 \\
- 2\frac{\mu}{N_{\ell}^2}\int_0^{ t} \int_0^{ t}e^{\mu(s_1+s_2-2 t)} \langle u_{ z_n}'(s_1) b_n'(s_2)\rangle_e \, ds_1ds_2.
\end{split}
\label{eq:var_noise_full}
\end{equation}
Using that $\langle u_{z_n}'(s_1) u_{z_n}'(s_2) \rangle_e = 2 D_u \delta(s_1-s_2)$ (with $\delta$ Dirac's delta function) and, similarly, $\langle  b_n'(s_1)  b_n'(s_2) \rangle_e = 2D_b \delta(s_1-s_2)$, as both $u_{z_n}'$ and $ b_n'$ are modeled as Gaussian white noise processes, the first two integrals in Eq.~(\ref{eq:var_noise_full}) give $D_u/\mu \left[1-\exp{(-2\mu t)}\right]$ and $D_b/\mu \left[1-\exp{(-2\mu t)}\right]$, respectively. As for the third term, involving the cross-correlation $\langle u_{ z_n}'(s_1) b_n'(s_2)\rangle_{\phi}$, numerical simulations suggest that it is negligible, and therefore we ignore it. The variance of $\xi_n$ then becomes
\begin{equation}
\langle \xi_n^2\rangle_e = \left( \frac{D_u}{\mu} + \frac{\mu D_b}{N_\ell^4} \right) \left( 1 - e^{-2\mu  t}\right).
\label{eq:var_noise}    
\end{equation}
Finally, plugging the results from Eqs.~(\ref{eq:cross-term}), (\ref{eq:var_trans_wave}), (\ref{eq:var_noise}) into the expression of the full variance of the particle displacement, one obtains the expression reported in Sec.~\ref{sec3},
\begin{equation}
\langle \xi^2 \rangle_e = \frac{A_w^2}{2 N_{\ell}^2}\left[ 1+e^{-2\mu t}-2 \, e^{-\mu t}\cos{(N_{\ell} t)}\right] + 
\left( \frac{D_u}{\mu}+ \frac{\mu D_b}{N_{\ell}^4}\right)\left( 1-e^{-2\mu t}\right) \, . 
\label{eq:var_full}
\end{equation}


%

\end{document}